\documentclass[conference]{IEEEtran}
\usepackage{cite}

\ifCLASSINFOpdf
  \usepackage[pdftex]{graphicx}
\else
\fi
\usepackage{amsmath}

\usepackage{todonotes}

\usepackage{amssymb}

\begin{document}
%
\title{Software-defined Radio Readout System for the ECHo Experiment}

\author{\IEEEauthorblockN{O. Sander, N. Karcher, O. Kr\"omer, M. Weber}
\IEEEauthorblockA{ Institute for Data Processing and Electronics (IPE) \\
Karlsruhe Institute of Technology (KIT)\\
Karlsruhe, Germany \\
Email: sander@kit.edu, karcher@kit.edu}
\and
\IEEEauthorblockN{S. Kempf, M. Wegner, C. Enss}
\IEEEauthorblockA{Kirchhoff-Institut for Physics\\	
	University of Heidelberg\\
Heidelberg, Germany\\
Email: sebastian.kempf@kip.uni-heidelberg.de}


%
}


\maketitle

\begin{abstract}
Metallic magnetic calorimeters (MMCs) are cryogenic detectors that offer an excellent energy resolution, a signal rise time of \textless100\,ns, a high dynamic range and almost optimal linearity. MMCs are of high interest for many experiments. One of them, the ECHo experiment, requires the utilization of large MMC detector arrays. The readout of such MMC arrays is a challenging task, which can be tackled using software-defined radios (SDRs). Though SDR is a well-known approach in communications engineering, a dedicated implementation for frequency division multiplexed readout of MMCs is new and one of the technological key elements of the ECHo project. ECHo will be the first experiment to use microwave SQUID multiplexed MMC detectors and therefore pioneering the hardware, firmware and software development in this domain. This paper presents the detailed concepts and current status of the development.
\end{abstract}


%
\IEEEpeerreviewmaketitle


\section{Introduction}

The evolution of physics experiments is often rendered possible by technical progress including increasing sensitivity of the detectors used to achieve higher energy resolutions, an increasing number of detectors for better spatial resolution or higher statistics. This is all based on the assumption that the resulting increase of data rate can be recorded and processed. 

The ECHo experiment is an excellent example of this type of experiments. Highly precise, novel cryogenic detectors, so-called magnetic metal calorimeters (MMCs), are used, which on the downside require a much more complex readout than conventional semiconductor detectors. In consequence, the development of a new type of readout electronics is needed, which in this case is based on the principle of a software-defined radio (SDR). This SDR uses a frequency division multiplexing (FDM) scheme, which allows the signals of several hundred detectors to be encoded on a common readout line. Thereby the number of lines between the detector array and readout electronics is heavily reduced, which is particularly advantageous when using cryogenic detectors such as the MMC. ECHo's SDR system consists of  mixing electronics to cover the frequency range between 4 and 8 GHz, a stage for the analog-digital or digital-analog conversion, as well as digital electronics whose task is the separation into the individual channels and the subsequent detection of events. 

Since an SDR readout system with the required bandwidth is not commercially available, an in-house development must be carried out within the framework of the ECHo experiment. This novel system is presented in this paper. An in-house development is also promising because the use of electronics is not limited to ECHo, but can also be used for other MMC-based experiments. This is particularly interesting because at the same time MMCs are a key technology for a large number of experiments and various domains including  high resolution X-ray spectroscopy\cite{Simmnacher2003}, nuclear forensics\cite{Okada2014}, radiation metrology\cite{Mitsuda2012} or direct neutrino mass investigation.

The remainder of this paper is structured as follows: Section~\ref{sec:fundamentals} will introduce the ECHo experiment, the MMC detector technology and the innovative frequency division multiplexing readout scheme. The following section presents the prinicple concept of using a software-defined radio for the ECHo readout. Section~\ref{sec:mixing_stage} presents the current status of the mixing stage. Section~\ref{sec:converter_mezzanine} introduces the converter mezzanine that is built using various high-speed converters. The following section~\ref{sec:digital_processing_hardware}
, gives an insight in the digital processing hardware as well as the signal processing chain implemented within the FPGA. Section~\ref{sec:conclusion} concludes the paper.

\section{Fundamentals}
\label{sec:fundamentals}

\subsection{The ECHo experiment}
The \textit{Electron Capture in  $^{163}$Ho experiment}(ECHo) is designed to reach sub-eV sensitivity on the electron neutrino mass by means of the analysis of the calorimetrically measured electron capture (EC) spectrum of the nuclide $^{163}$Ho. This radioactive isotope decays to Dysprosium with emission of an Electron-Neutrino:
\begin{figure}[t]
	\centering
	\includegraphics[width=0.9\columnwidth]{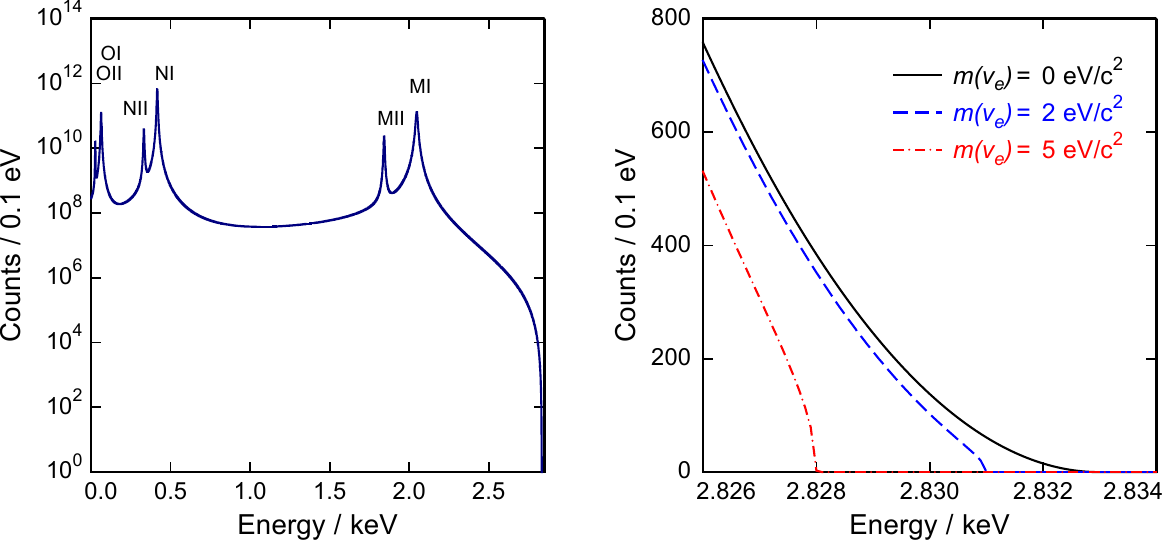}
	\caption{Left: $E_{EC}$ spectrum of the $^{163}\mathrm{Ho}$ decay with $10^{14}$ events. Right:  ECHo endpoint energy plot for different Electron-Neutrino energies.\cite{Gastaldo2017}}
	\label{fig:echo-spectrum}
\end{figure}

\begin{eqnarray}
	^{163}\mathrm{Ho} + e^- \rightarrow\  ^{163}\mathrm{Dy}^{*} + \nu_e\\
	^{163}\mathrm{Dy}^{*} \rightarrow\  ^{163}\mathrm{Dy}+E_{EC} 
	\label{math:hodecay}
\end{eqnarray}

The maximum energy available for this decay $Q_{EC}$, approx. 2.833\,keV, can be measured using high-precision Pennig-trap mass spectroscopy\cite{Gastaldo2017}. However, the direct measurement of the emitted Neutrino is not feasible. Though the measurable energy $E_{EC}$ emitted during decay reflects the mass difference $Q_{EC}$ without the mass of the Neutrino. The released $E_{EC}$ (see Fig. \ref{fig:echo-spectrum}) forms a continuous energy spectrum with only a few events in the interesting endpoint region.

The energy spectrum of this decay determines the kind of detectors to be used. For ECHo arrays of low temperature metallic magnetic calorimeters (MMCs) are being developed to measure the $^{163}$Ho EC spectrum with an energy resolution below 3\,eV full width half maximum. The detectors have a time resolution below one $\mu$s. Increasing the amount of events in the endpoint region is essential thus measurement requires the acquisition of a high-resolution and high-statistics spectrum with up to $10^{14}$ decays. Due to constraints on the activity per pixel to limit internal background caused by a pile-up of detector events, large detector arrays with up to $10^6$ detectors running in parallel are required to acquire such a high statistics spectrum in finite time for the next stage of the experiment. Further details on the experiment can be found in \cite{Gastaldo2014} and \cite{Gastaldo2017} and are therefore not presented here.


\subsection{Metallic Magnetic Calorimeters}

\begin{figure}[b]
\centering
		\includegraphics[width=0.8\columnwidth]{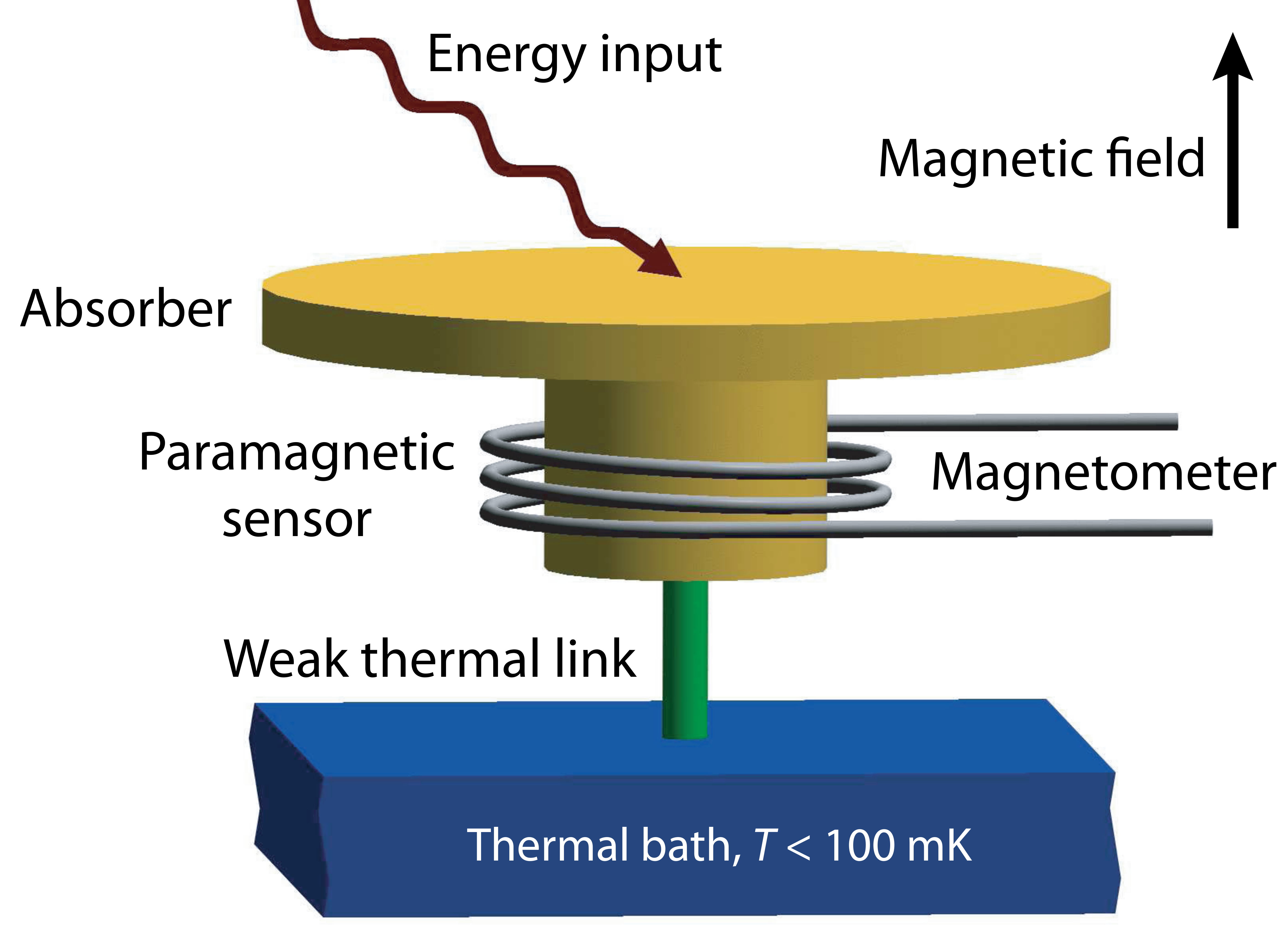}
		\caption{Schematic drawing of a metallic magnetic calorimeter \cite{Gastaldo2009}}
		\label{fig:mmc_schema}
\end{figure}

Metallic magnetic calorimeters are typically operated at temperatures well below 100\,mK and make use of a metallic, paramagnetic temperature sensor to transduce the temperature rise of the detector upon the absorption of an energetic particle into a change of magnetic flux (see Fig.~\ref{fig:mmc_schema}). The latter can be precisely measured using a superconducting quantum interference device (SQUID). This outstanding interplay between a high-sensitivity magnetic thermometer and a near quantum-limited amplifier results in a very fast signal rise time reaching values below 100\,ns. Furthermore, they achieve an excellent energy resolution which is competitive to the resolving power of wavelength-dispersive crystal spectrometers, a large energy dynamic range, a high quantum efficiency as well as an almost ideal linear detector response. For this reason, a growing number of groups located all over the world are developing MMC arrays of various sizes which range from a few to several thousand pixels. These arrays are routinely used in a variety of applications and often appear to be a key technology for measurements that require high-resolution and wide-band energy-resolving detectors.  
MMCs have already been proven to be eminently suited detectors for various applications such as high resolution X-ray spectroscopy\cite{Simmnacher2003}, nuclear forensics\cite{Okada2014}, radiation metrology\cite{Mitsuda2012} or direct neutrino mass investigation.

\subsection{Microwave SQUID Multiplexing}

\begin{figure}[t]
	\centering
	\includegraphics[width=0.8\columnwidth]{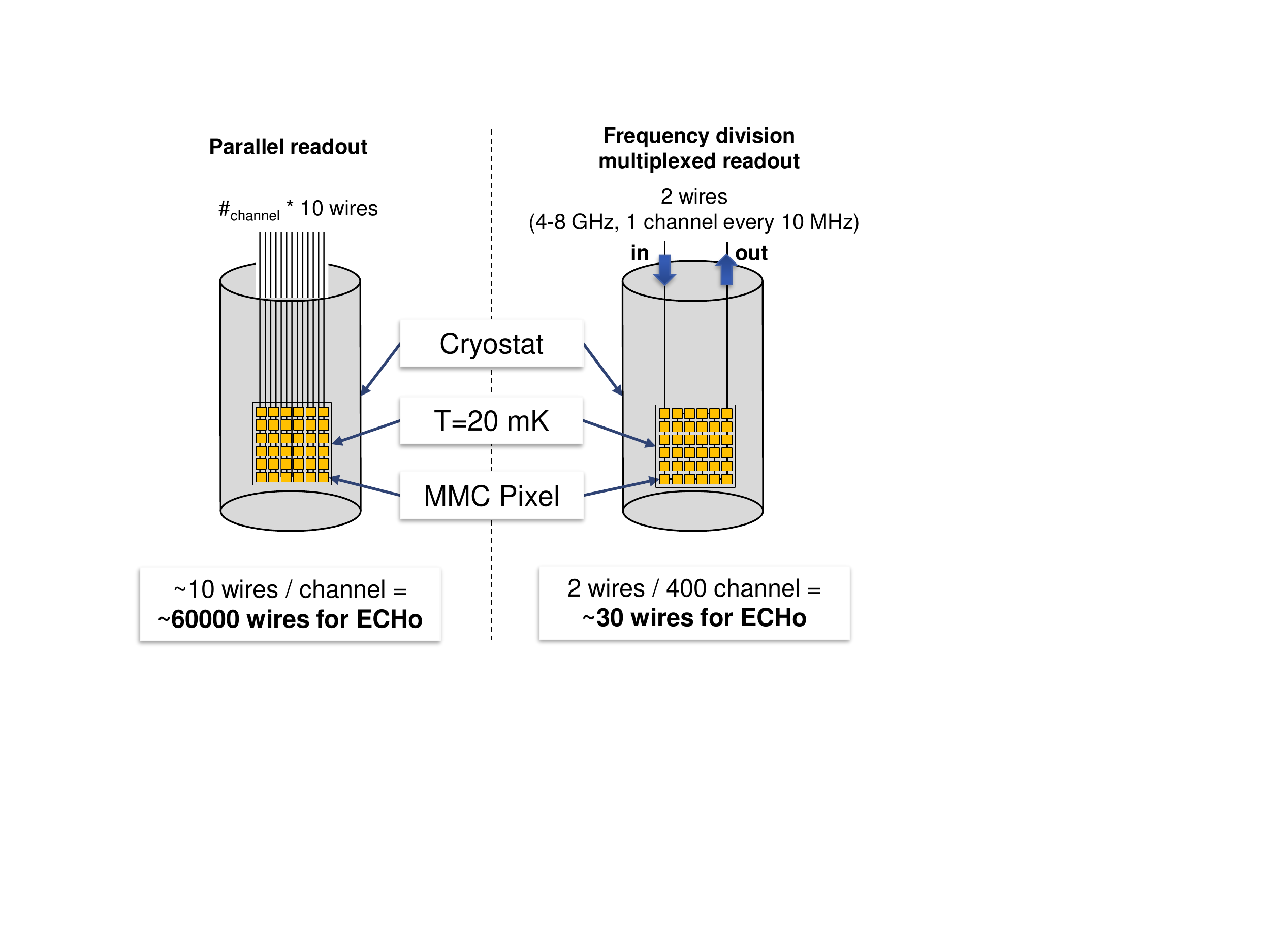}
	\caption{Microwave SQUID multiplexing versus dedicated readout lines}
	\label{fig:parallel_vs_fdm}
\end{figure}

\begin{figure}[b]
	\centering
	\includegraphics[width=0.9\columnwidth]{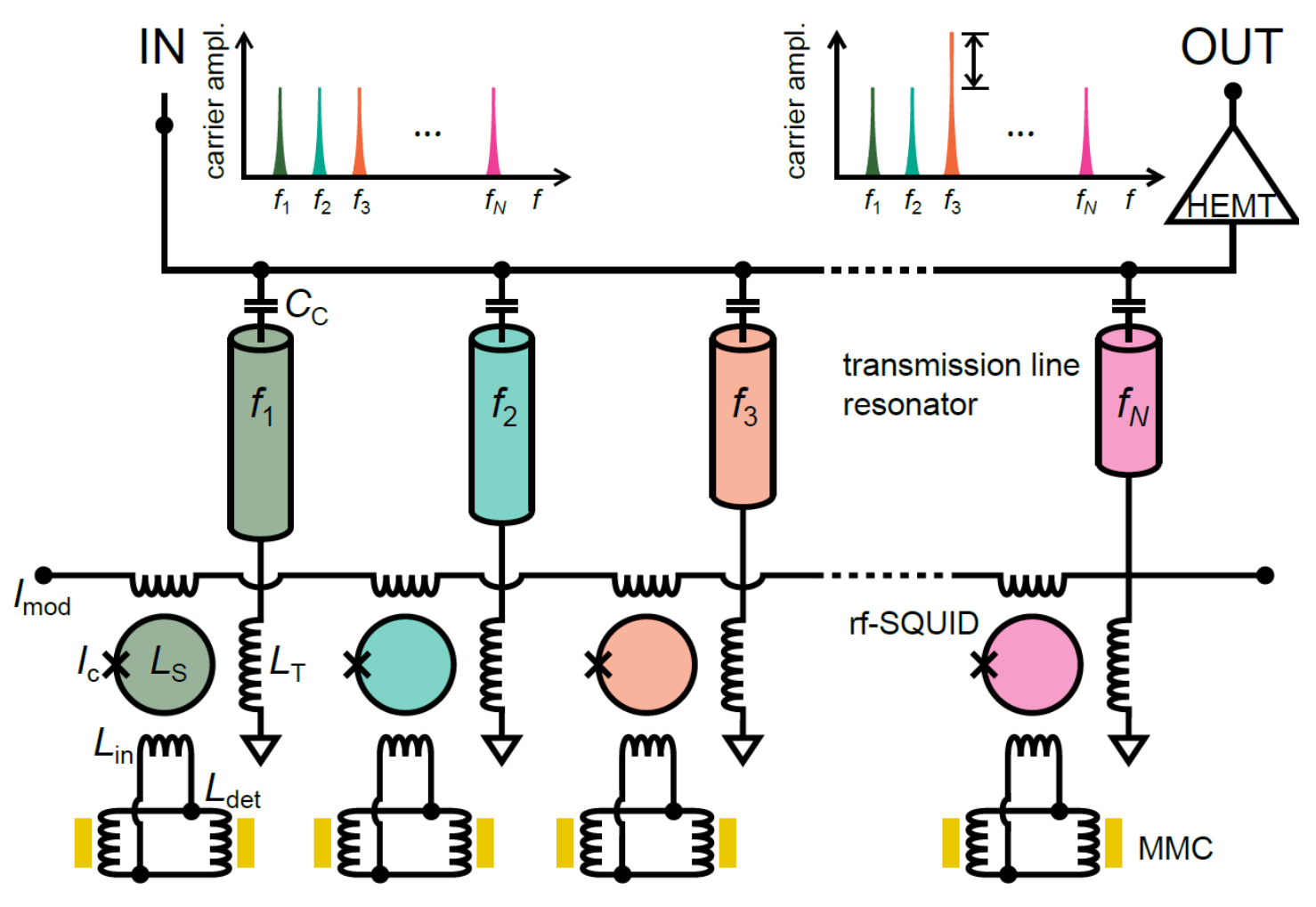}
	\caption{Microwave SQUID multiplexing architecture: An array of transmission-line coupled resonators with individual $f_r$ is used for multiplex. Each resonator is tuned by the inductance change of the rf-SQUIDs due to a magnetic flux from the MMC. There are two sensor pixels per resonator: one tunes to higher frequencies, the other to lower.}
\end{figure}

An efficient readout of large MMC arrays can be achieved through microwave SQUID multiplexing \cite{Mates2008, Kempf2017a, Kempf2017b} that allows to connect a large number of pixels to one single readout line (Fig. ~\ref{fig:parallel_vs_fdm}). The number of wires leading into the cryostat can be reduced by 99.95\,\% .

It requires three different components or techniques for proper operation: (1) The actual cryogenic multiplexer which modulates the detector signals onto different carrier signals and combines them into a single readout line, (2) a room temperature readout electronics for operating the multiplexer as well as (3) a method for linearizing the multiplexer response to maintain the intrinsic linearity of MMCs. 

Each channel of a microwave SQUID multiplexer consists of a non-hysteretic, unshunted, current-sensing rf-SQUID \cite{Mates2008}.
The SQUID is inductively coupled to a load inductor terminating a superconducting $\lambda/4$-transmission line resonator having a resonance frequency in the one-digit GHz range. 
Due to the mutual interaction between the SQUID and the load inductor, the circuit's resonance frequency gets magnetic flux dependent and is shifted as the magnetic flux introduced to the SQUID loop changes. For simultaneous readout of $N$ detectors, $N$ resonance circuits, each with a unique resonance frequency, are capacitively coupled to a common transmission line. This arrangement allows the measurement of the actual resonance frequencies of all channels by injecting a microwave frequency comb into the common transmission line. All individual comb tones are set with respect to the undisturbed resonance frequency of the different channels. The sensor signals modulate amplitude or phase of each transmitted tone of the frequency comb by using standard homodyne or heterodyne detection techniques.

\begin{figure}[h]
\centering
\includegraphics[width=0.75\columnwidth]{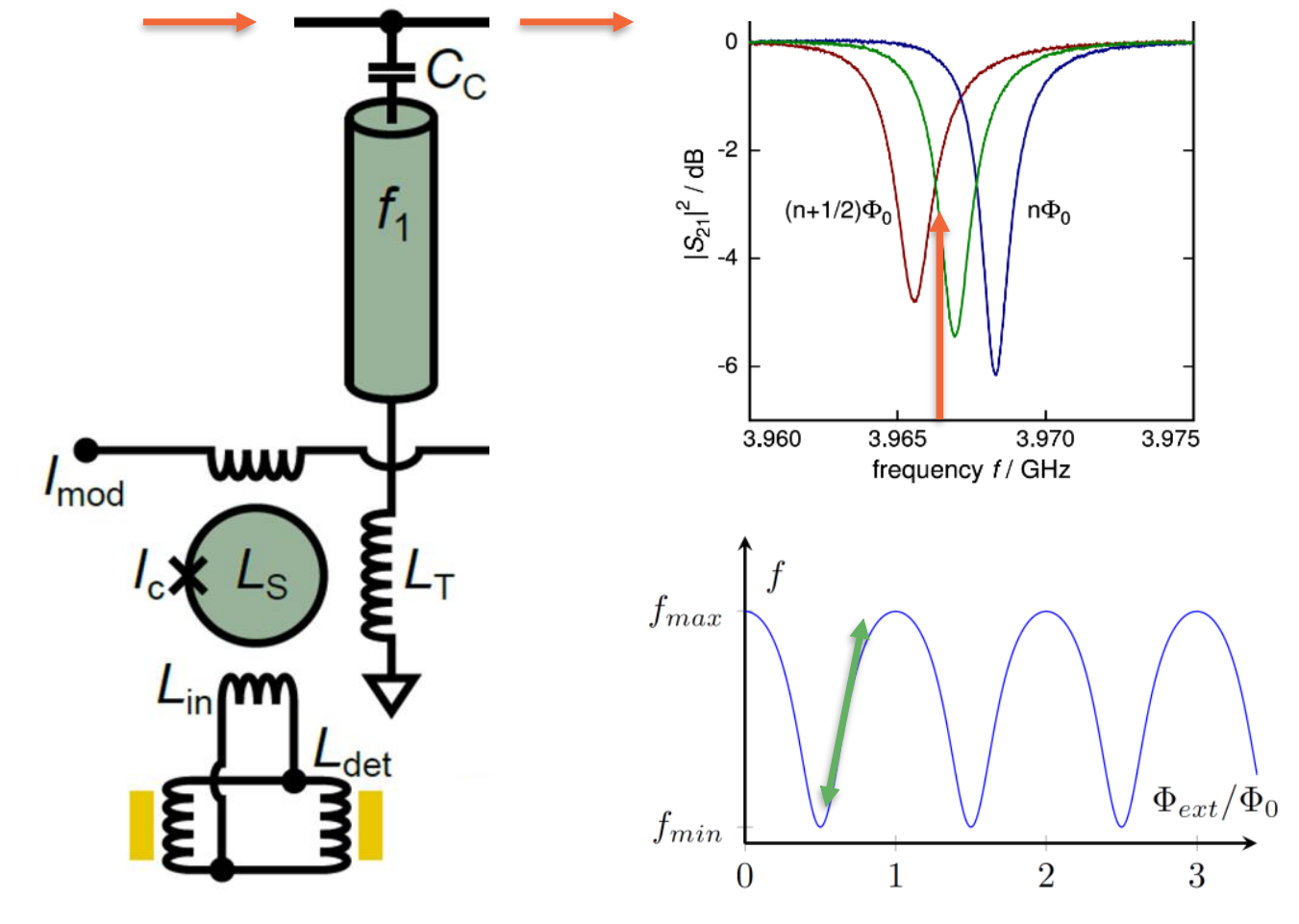}
\caption{Left: Resonator coupled to transmission line. Orange arrows marks stimulation signal. Top, right: $S_{21}$ parameter, that is changing according to magnetic flux. Stimulation signal is tuned to the steepest point of the resonance dip. Bottom, right: Theoretical tuning curve frequency over magnetic flux. This curve is periodic over the flux quantum. Highest sensitivity is also reached at steepest point.}
\label{fig:tuning}
\end{figure}

\begin{figure}[b]
	\centering
	\includegraphics[width=0.9\columnwidth]{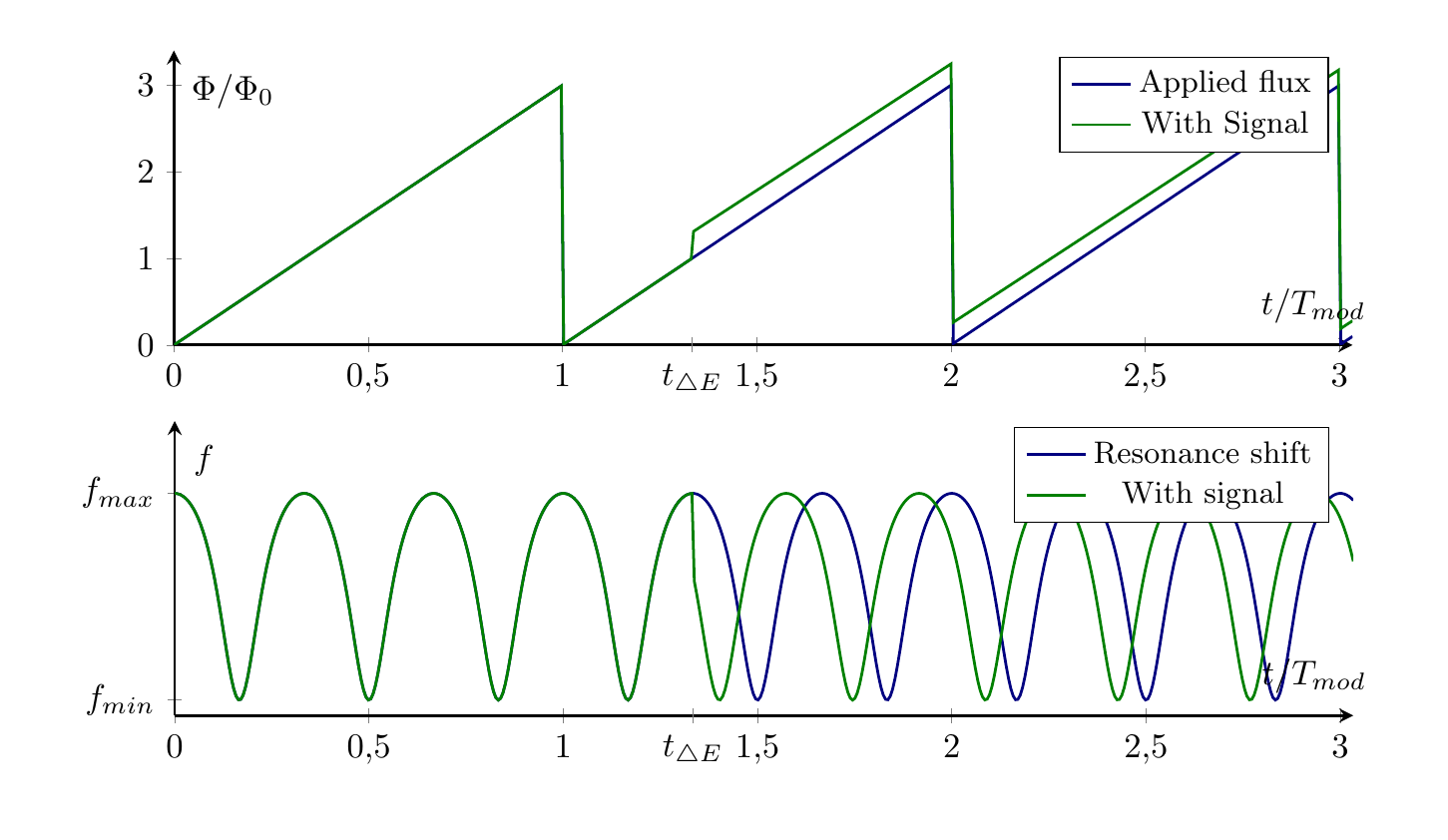}
	\caption{Top: linear flux-ramp applied as $I_\mathrm{mod}(t)$ lead to a flux $\Phi(t)$ . Bottom: Resulting resonance frequency shift. Figure shows a comparison between idle signal and shifted signal due to energy disposition.}
	\label{fig:fluxramp}
\end{figure}

This approach assumes the rf-SQUID is tuned to the steepest operating point of the flux tuning curve (Fig. \ref{fig:tuning}) to get a large resonance shift on a small flux change. Since the flux offset in this curve is arbitrary for each pixel and large flux signals of events lead to a non-linear working condition an external tuning of the SQUID is required. This additional individual biasing would make the advantage of the frequency multiplex void. However, the signal for tuning can be modulated by linearly ramping the $I_{mod}$  (Fig. \ref{fig:tuning}). The resulting frequency shift will cause a sinusoidal amplitude modulation. Any flux offset due to an event will lead to a phase shift of the modulation (Fig. \ref{fig:fluxramp}). The information about measured events now lies in the phase of the this signal. This phase can be demodulated to reconstruct the sensor events.\cite{Mates2012}

\section{Software-defined Radio Approach}
\begin{figure}[t]
	\centering
	\includegraphics[width=0.9\columnwidth]{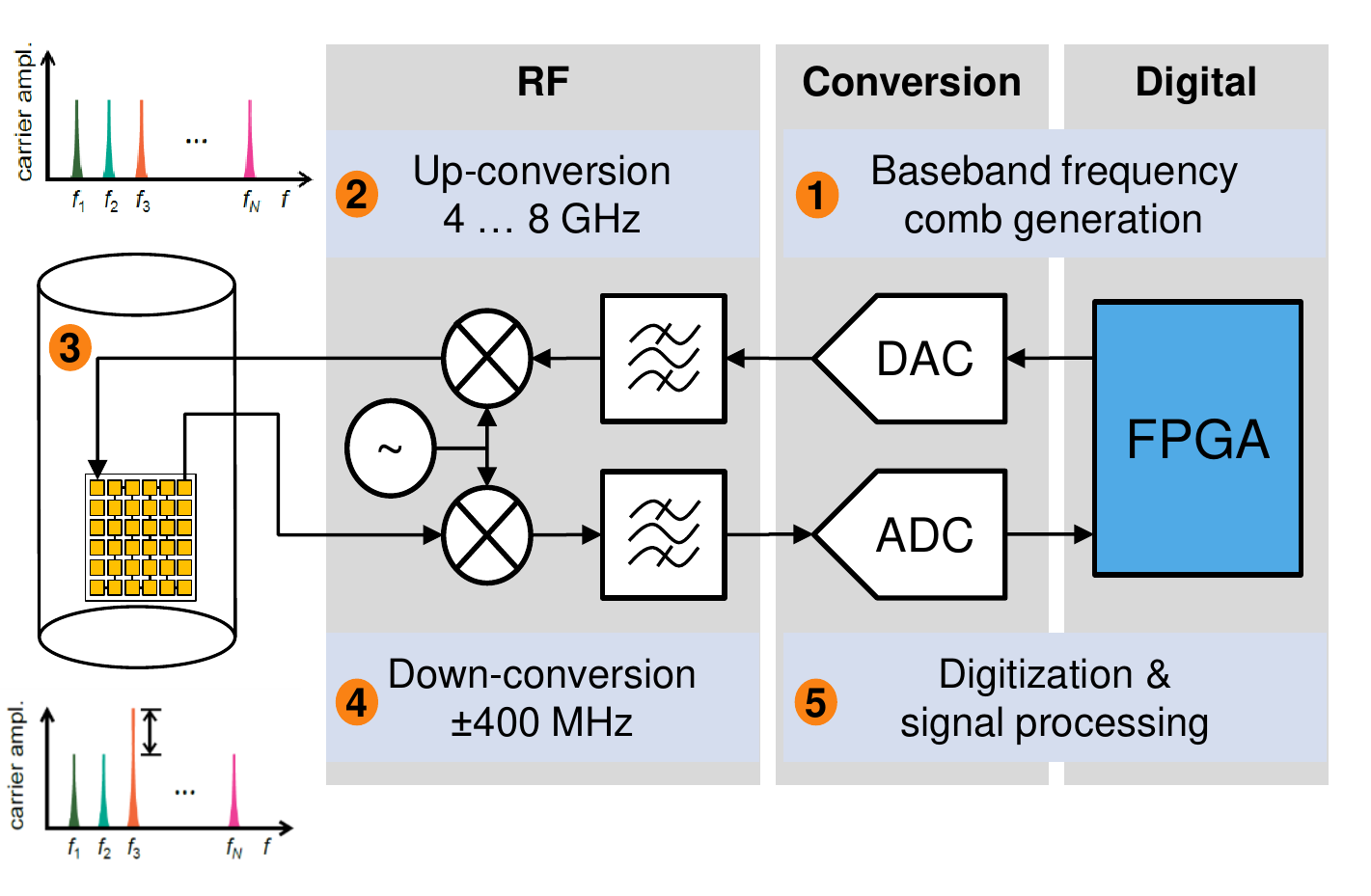}
	\caption{Software Defined Radio (SDR) Principle Approach}
	\label{fig:sdrconcept}
\end{figure}
A straightforward technique to generate a microwave frequency comb as well as to determine the amplitude or phase of each transmitted carrier signal is software-defined radio (SDR). As shown in Figure~\ref{fig:sdrconcept}, the SDR architecture can be separated into three sections: RF, digitization and digital processing.

The digital electronics generate the frequency comb in the baseband with a bandwidth of several 100\,MHz being further processed by two fast digital-to-analog converters (DACs). The RF electronics then up-converts the frequency comb into the frequency band of the SQUID multiplexer using an \textit{I/Q} mixer to match the resonance frequencies of the resonators. After injecting the frequency comb into the cryostat and passing the multiplexer, the modulated frequency comb is mixed by the RF electronics into the baseband again using an \textit{I/Q} down-mixer. Both mixers are driven by the same local oscillator signal. The resulting in-phase and quadrature baseband signals are digitized by two fast analog-to-digital converters (ADCs). The subsequent signal processing, i.e. the separation of the frequency-comb into the different carrier signals, the determination of amplitude and phase of each carrier signal as well as the determination of the characteristic parameters of each detector signal, is then performed in real-time by using a field programmable gate array (FPGA).

The digital electronics are based on a heterogeneous Xilinx Zynq Ultrascale+ FPGA \cite{XilinxUSP}, which is a combination of a Kintex Ultrascale+ FPGA fabric with a powerful ARM Cortex-A53 based quad-core processor system along with two ARM Cortex-R5 cores. The processors perform slow control including calibration and data transmission to the storage backend. All significant processing steps are executed in a full custom signal processing chain in hardware within the FPGA part of the system. The frequency-comb generator stores and plays precomputed waveforms and therefore is a simple arbitrary waveform generator. The MMC detectors also require a flux-ramp, which is also generated in the FPGA by a specific hardware module. On receiver site, the samples coming from the ADC are at first processed in a channelizer, which extracts the various channels. Afterwards, the flux-ramp is demodulated for each channel. Then the detector signals are available and the relevant parameters can be obtained in the last stage. After all relevant event parameters were extracted, the data is handed over to the software layers. These pack the data and transmit it to the backend storage system. 

The next stage of the ECHo experiment is planned to achieve 100\,kBq. The activity limit per pixel is assumed to be at  10\,Bq per pixel, which results in a number of 10\,k pixels. Due to the use of double pixels per channel 5 k readout channels are required. Furthermore a channel spacing of 10 MHz is planned along with a usable bandwidth between 4 GHz and 8\,GHz per readout system. In other words, each readout system will be able to handle 400 channels. Eventually this means ECHo will require at least 13 readout systems to achieve its goal. In order to have some safety margin, the experiment decided to deploy 15 readout systems, therefore being potentially able to handle 6\,k channels resulting on 120\,kBq of activity. 


\section{Wide-band mixing stage}
\label{sec:mixing_stage}

The microwave resonators are located in a band between 4 to 8\,GHz, mostly for two reasons. First the cryogenic setup contains a HEMT amplifier, which has a limited frequency bandwidth, and second the physical length restrictions of the line resonator on the chip sets the usable resonance frequencies to this 4 to 8\,GHz range. Direct sampling is not possible, because high SNR digital-to-analog and analog-to-digital converters are limited to much lower frequencies $<4$\,GHz. Therefore an up- and down-conversion between electronics in the cryostat and the converters is required. Covering a 4 GHz bandwidth also implies the use of multiple ADCs and DACs because of the limited sampling rate and nyquist criteria. Thus the RF electronics also merge the sources (DAC to RF) and splits the incoming signal into various channels (RF to ADC). 

To cover the full spectrum several double sideband mixers are used. Each mixer has 800\,MHz complex bandwidth, mixed to $4.4 + 0.8\cdot n \,\mathrm{GHz}$ and then combined to a common spectrum. On receiver side several double sideband mixers are used for down-mixing of 800 MHz band snippets into complex baseband signals. 

Since each up- and down-mixing pair requires a unique carrier frequency, cost efficient phase-locked loop chips with integrated voltage controlled oscillators will be used on the RF board. Such as a low phase noise LMX2594 from Texas Instruments can be chosen, which covers frequencies up to 15\,GHz. Selection and characterization of other components is still an ongoing process and scheduled this year.

\begin{figure}[htbp]
\centering
\includegraphics[width=1\columnwidth]{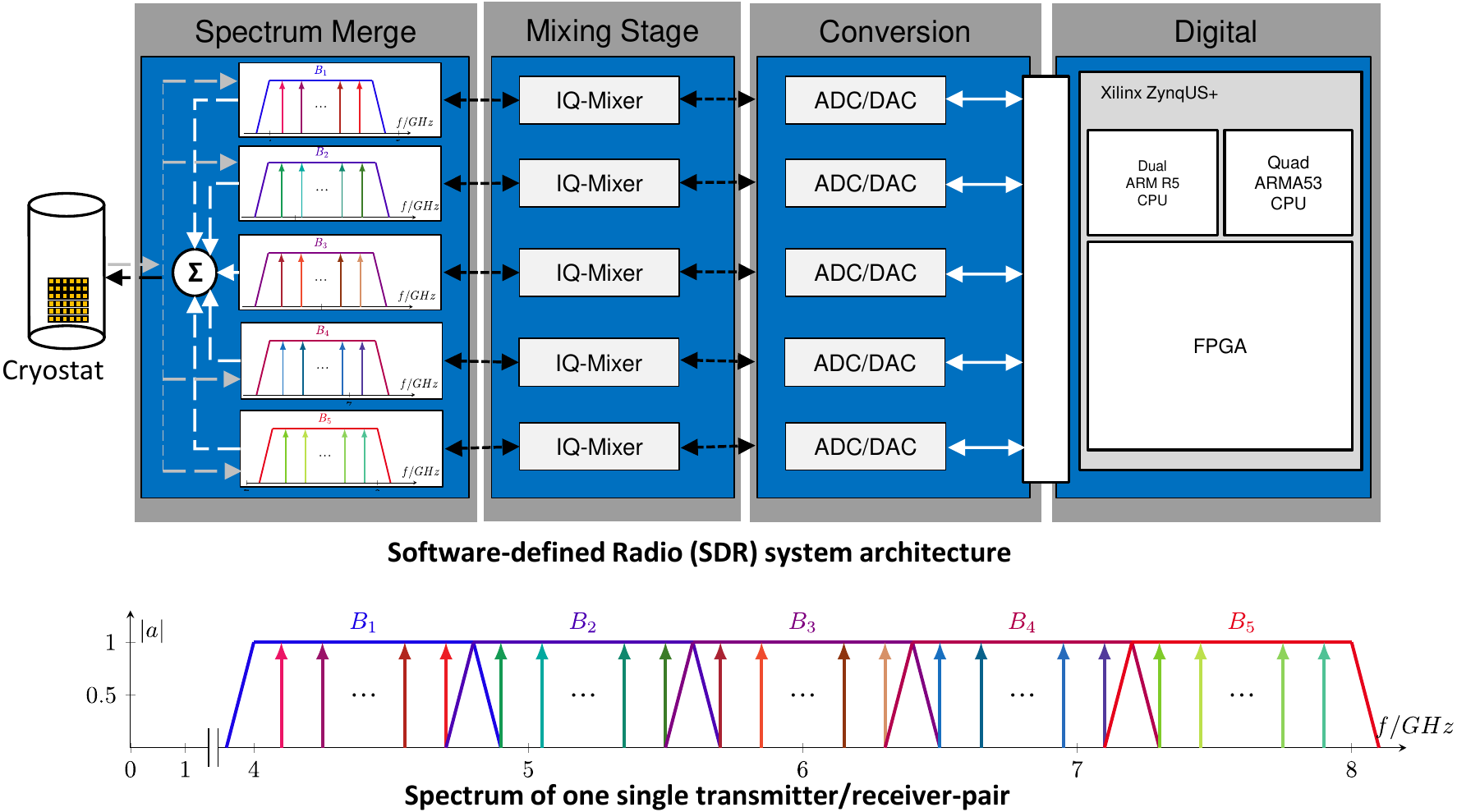}
\caption{Wide-band Mixing Stage Architecture}
\end{figure}


\section{Modular converter mezzanine}
\label{sec:converter_mezzanine}
\begin{figure}
	\centering
	\includegraphics[width=0.9\columnwidth]{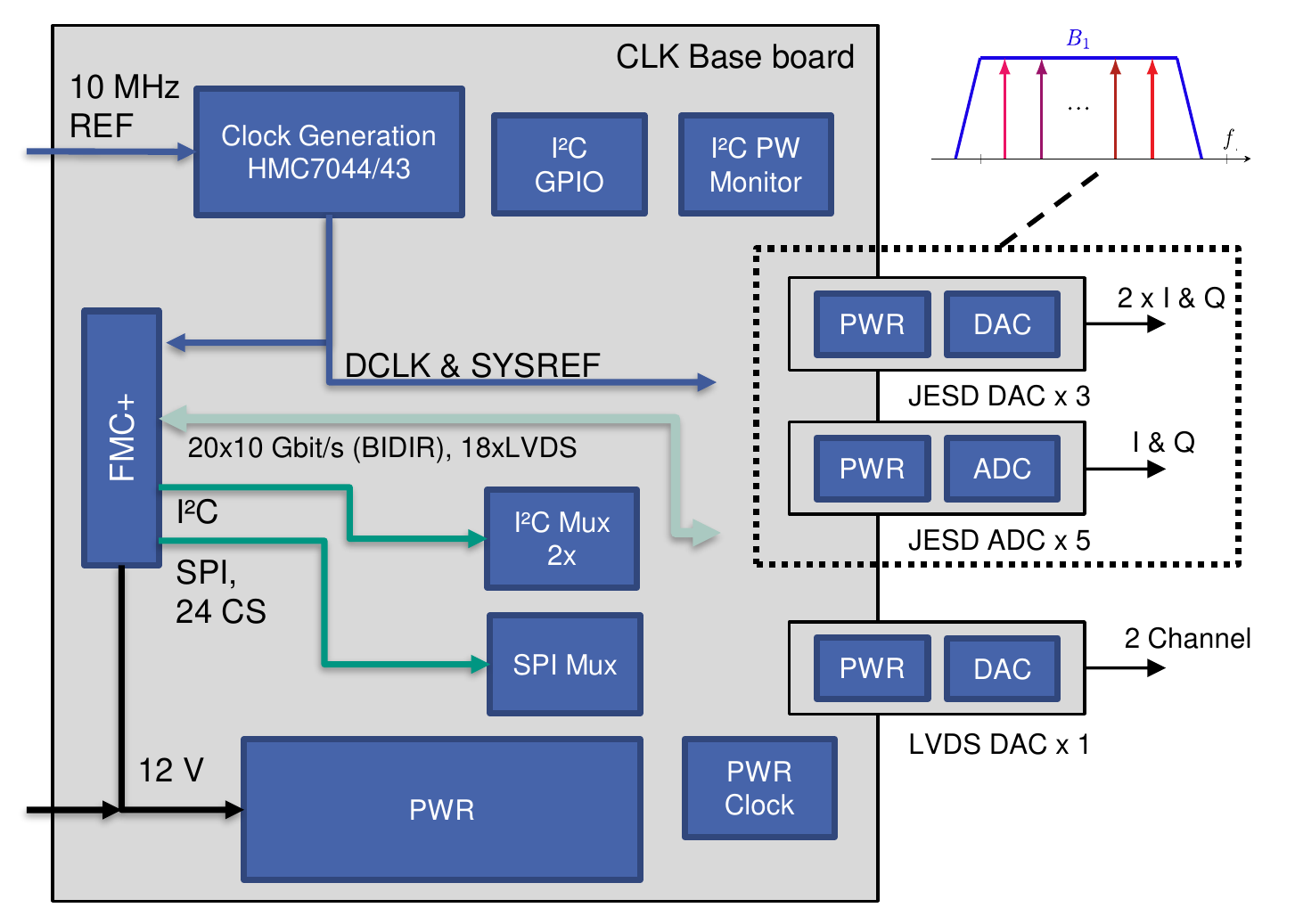}
	\caption{Modular Converter Architecture}
\end{figure}
The intended bandwidth of 4\,GHz require at least four dual channel DAC and ADC ports. However, considering the aliasing filters with the passband of 400\,MHz, the total amount increases to five pairs. 

The mezzanine is an assembly of one clock baseboard, five AD-conversion modules, three DA-conversions modules and one additional slow DA-conversion module. The baseboard is aligned with the VITA57.4 draft and offers a FMC+ connector for connecting the FPGA board. In comparison to the regular FMC connector, the FMC+ offers four additional banks with 16 multi gigabit datapaths and four transceiver clock signals. Only a total of 20 datapaths is required for the converters. It is intended to offer an FMC-only mode to be backwards compatible with last generation FPGA board. However this supports only a subset of one DAC and two ADC daughter boards. The LVDS signals of the FMC connector hold SPI, $\mathrm{I^2C}$, interrupt and status signals. Also the slow DAC receives the data via 18 LVDS lines of the FMC+ connector.

Offering a low jitter clock to maintain the high SNR of the converters is crucial. Therefore a HMC7044 two stage jitter-cleaning PLL clock chip has been chosen. The first stage locks an external 100 MHz VCXO to a reference clock, the second stage locks an internal 3 GHz VCO to the VCXO. In simulation the cascaded setup showed an total of 115\,fs. The maximum $\mathrm{SNR}_\mathrm{jitter}$ at 400 Mhz is 70.7\,dB, therefore higher than the intrinsic SNR of the ADC. It can be calculated by the following formula:

\begin{equation}
\mathrm{SNR}_\mathrm{jitter} = -20\log_{10}(2\pi\cdot f_{in}\cdot T_{jitter})
\label{math:jitter}
\end{equation}

For analog-to-digital conversion an AD9680 two channel 1\,GS/s type was chosen. The converter offers a good SNR of over 65\,dB. Furthermore it offers four on-chip digital down conversions. This enables a preprocessing from two to four sub-bands covering a bandwidth of 780\,MHz. With the applied decimation a parallel sample processing can be avoided. The ADC is connected trough the FMC+ connector with four 10\,Gbit/s high speed links and uses the JESD204b protocol for data transmission. The corresponding DAC is a AD9144 type with four channels and 2.8\,GS/s conversion speed. It will be operated with a data rate of 1\,GS/s. One of the three DAC modules is used in two channel mode only. The connection protocol is an eight/four lane JESD204b operated at 10\,Gbit/s. 

The ac-coupling of the high bandwidth converters is designed with a double balun configuration. It enables a total bandwidth of 4.5\,MHz to 3\,GHz. Neither aliasing filters nor reconstruction filters are integrated on the PCB.

To generate the required flux ramp for the sensors a second DAC module provides a low distortion, low noise current. As DAC a two channel 500\,MS/s MAX5898 is used. The design bandwidth of the DC-coupled analog electronics is below 30\,MHz. The coupling is achieved with a double operational amplifier circuit with two ADA4899. The circuit converts from differential to single-ended and offers a signal offset by adding the output of a AD5697R $\mathrm{I^2C}$ DA converter. The module can either generate a voltage or act as a voltage controlled current source.

\section{Digital processing hardware}
\label{sec:digital_processing_hardware}

\begin{figure}[b]
	\centering
	\includegraphics[width=0.90\columnwidth]{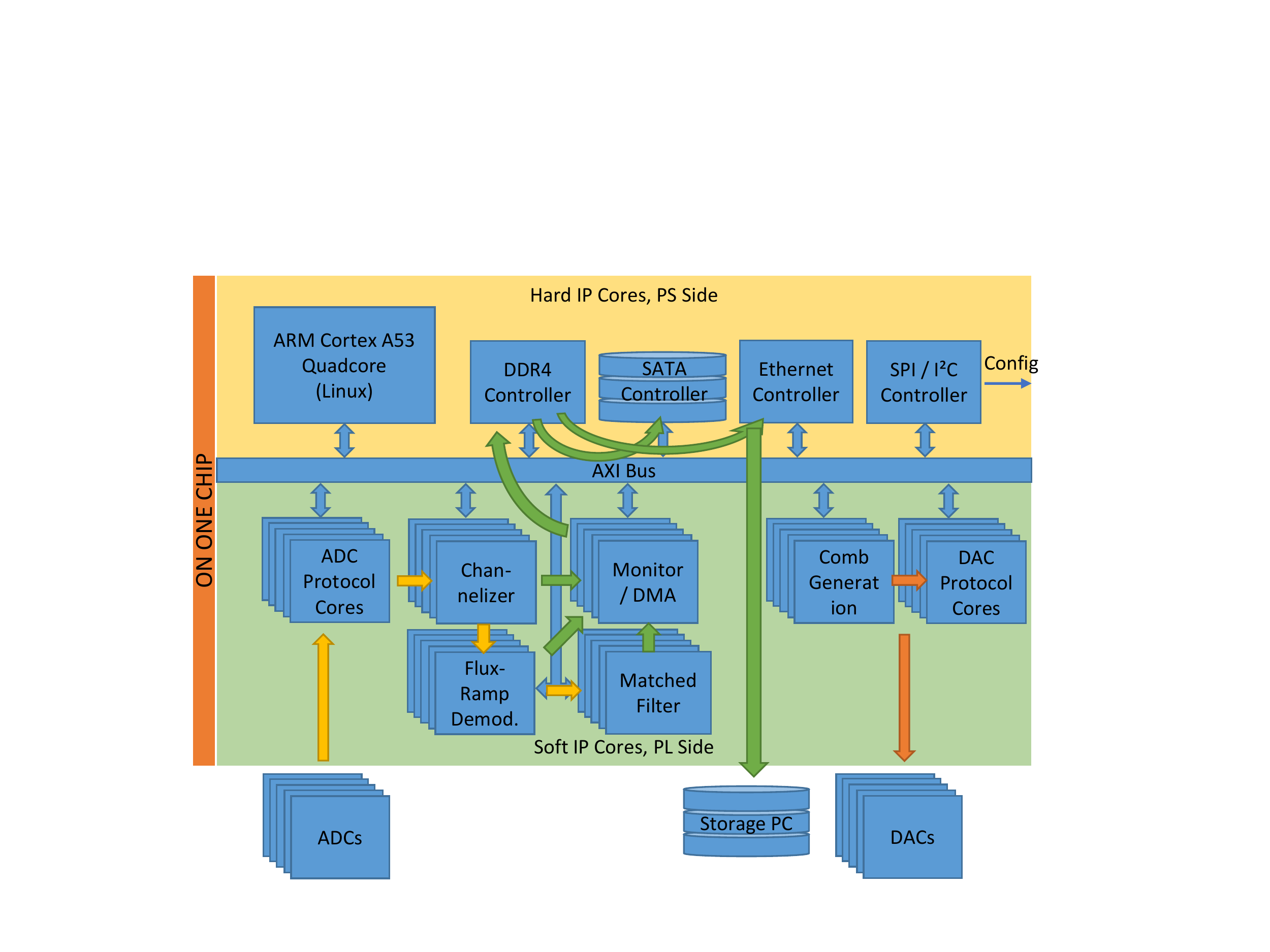}
	\caption{Architecture of the FPGA firmware}
	\label{fig:fpga-firmware}
\end{figure}

The processing and controlling backend will be a new, second generation of the IPE HiFlex board (last gen. \cite{1748-0221-12-03-C03015}) with a Xilinx Zynq UltraScale+ 11EG FPGA-SoC. This offers 2928 digital signal processing blocks and a 20 lane high speed interface through FMC+. A second board for processing can be connected via a Samtec FireFly link.

The Firmware concept is shown in Fig. \ref{fig:fpga-firmware}. Both ADC and DAC are connected via Xilinx JESD204b IP-cores to the signal processing cores. The frequency comb is preprocessed on the CPU side using a inverse FFT. The time domain samples are written in a cyclic buffer which is repetitively transferred to the DAC.

After gathering the sub-bands from the ADCs a further processing is required to split the individual resonator channels. Due to the intrinsic rise time of 100\,ns of the sensor, the signal bandwidth is limited to 1.6\,MHz. Within a spacing of at least 8.4\,MHz the stop band should be reached to suppress crosstalk. Furthermore a maximum decimation leads to the most efficient post processing. Factors of $2^n$ with $n = 64 \vee 128$ for decimation with a resulting sample-rate of 3.90625 or 7.8125\,MHz are suitable. 

During the channelization process, the signal is mixed with the complex sine of a numerical controlled oscillator. In combination with a low pass filter it demodulates the amplitude of the signal. Afterwards a module for flux-ramp demodulation converts a phase modulated block to one phase sample\cite{Mates2012}. The required sine wave for the demodulation can be shared amongst all channels. Since the flux-ramp is generated by the converter board any calculation is fully synchronized with the modulation. The resulting phase signal contains the pulses due to particle events. Particle's energy can be extracted from pulse height and pulse integral. In this connection a matched filter approach can be used\cite{Fleischmann2005}. After calibrating the sensor by a known energy value, the resulting correlation values can be scaled to the equivalent particle's energy.

All energies of events measured in the ECHo experiment should be stored in a central data center. The native support of the TCP/IP and 1\,GBit/s Ethernet connection of the Zynq FPGA is a major advantage as it reliably transfers the data trough network. Additionally it could provide a mirrored storage attached to SATA.  With 10\,Bq pixel activity, 64 bit per event and 800 pixel per device, a data rate of 62.5\,KB/s per device will be transferred. In case of a raw data logging of highest energy events ($1/10^5$-fraction) additional 2.56\,KB/s are generated, for $f_s=7.8125$\,MHz, 4 Byte sample size and 1\,ms event length. With 15 devices a total data rate of 975.9\,KB/s will be transferred.


\section{Summary}
\label{sec:conclusion}
The ECHo experiment requires the utilization of large MMC detector arrays. The readout of such MMC arrays is a challenging task, which can be tackled using software-defined radios (SDRs) as presented in the previous sections. Though SDR is a well-known approach in communications engineering, a dedicated implementation for frequency division multiplexed readout of MMCs is new and one of the technological key elements of the ECHo project. Each of the SDR systems has to process a relatively high input data rate in the range of 2.4~Tbit/s for 400 detector channels. After all online processing steps, the data rate is massively reduced to less than 8\,Mbit/s to be stored in the backend. This is only feasible by utilizing the massive parallelism offered by modern FPGAs.  ECHo will be the first experiment to use microwave SQUID multiplexed MMC detectors and therefore pioneering the hardware, firmware and software development in this domain. The current status has been presented in this paper. The next step will be the characterization of the final electronics which are currently in development.

\bibliographystyle{unsrt}
\bibliography{bibliography}

\end{document}